\documentclass[%
amsmath,
reprint,
amsmath,amssymb,
aps,
]{revtex4-1}

\usepackage[T1]{fontenc}
\usepackage{textcomp}
\usepackage{txfonts}
\usepackage{bm}
\usepackage{graphicx}
\usepackage{dcolumn}
\usepackage{bm}
\usepackage{xcolor}
\usepackage{amsmath}
\usepackage{lipsum} 
\usepackage{mathtools}
\usepackage{sidecap}
\usepackage{booktabs}
\usepackage{graphicx}
\usepackage{float}
\usepackage{anyfontsize}
\usepackage{hyperref}
\usepackage[utf8]{inputenc}
\usepackage{textgreek}

\hypersetup{colorlinks=true, linkcolor=blue, citecolor=blue, urlcolor=blue}

\begin{document}

\preprint{APS/123-QED}

\title{Quantum Fluctuations Drive Angular Momenta in Nuclear Fission}

\author{M. H. Zhou$^{1}$}
\author{S. Y. Chen$^{1}$}\thanks{S. Y. Chen's contribution is equal to M. H. Zhou's in this work and should be considered as co-first authors}
\author{Z. Y. Li$^{2,1}$}
\author{M. S. Smith$^{3}$}
\author{Z. P. Li$^{1}$}
\email{E-mail: zpliphy@swu.edu.cn}
\affiliation{%
$^{1}$School of Physical Science and Technology, Southwest University, Chongqing 400715, China\\
$^{2}$China Nuclear Data Center, China Institute of Atomic Energy, Beijing 102413, China\\
$^{3}$Physics Division, Oak Ridge National Laboratory, Oak Ridge, Tennessee, 37831-6354, USA}

\date{\today}

\begin{abstract}
\fontsize{10pt}{12.6pt}\selectfont
Quantum fluctuations are ubiquitous and play crucial roles across various scales and systems, such as the Big Bang, black hole dynamics, quantum phase transitions in microscopic many-body systems, and so on. Nuclear fission manifests as a complex nuclear shape stretching until it splits into fragments with substantial angular momenta, also exhibiting complex quantum fluctuations and specifically shape fluctuations. For over 40 years, researchers have puzzled how the fission fragment angular momenta are generated dynamically from (almost) zero spin, as well as the particular role played by quantum fluctuations. Here, for the first time, we report the quantum shape fluctuations that drive fragment angular momenta during nuclear fission, based on a global, microscopic, and dynamical simulation. The calculated probability distributions of fragment angular momenta are in good agreement with the experimental measurements, and the sawtooth-like mass dependence of average angular momenta is reproduced very well. It is noteworthy to find that the shape fluctuations---multiple rotations, vibrations, and their couplings---drive the generation and chaotic evolution of fragment angular momenta during fission fragment formation and induce strong correlations between angular momentum orientations of partner fragments at small, medium, and large opening angles ($\phi_{LH}\approx 30^\circ, 90^\circ, 160^\circ$). Our work not only deepens the fundamental understanding of the nuclear fission mechanism but also has implications for the $\gamma$-ray heating problem in nuclear reactors and the synthesis of superheavy elements.
\end{abstract}

\maketitle

\fontsize{11pt}{13.2pt}\selectfont
Quantum fluctuations are ubiquitous throughout the realm of physics, influencing phenomena from the cosmic scale to the subatomic. They are instrumental in the inflationary model of the Big Bang \cite{ghosh1998PLB,Bojowald2008PRL,GreenPRL2020,GreenPRL2020,Scott1006Science} and the formation of primordial black holes \cite{Saito2009PRL,CaiYiFu2018PRL}. Additionally, these fluctuations drive quantum phase transitions \cite{si2001nature,greiner2002Nature,osterloh2002Nature,Saito2018NC,Puschmann2020PRL,Ivanov2020PRL,Nandi2022PRL,li2022Nature}, enabling a system to change states without thermal energy. Similarly, the atomic nucleus, as a strongly correlated quantum many-body system, also demonstrates significant quantum fluctuations \cite{Bohrbook}. In particular, quantum shape fluctuations are evident in the fission of heavy nuclei, where the nuclear shape undergoes complex stretching and evolves into two (or more) fragments with substantial angular momenta. For over 40 years, researchers have strived to understand how the fission fragment angular momenta are dynamically generated from (almost) zero spin, as well as the specific role of quantum fluctuations in this critical process \cite{Schmidt2018}. To this end, significant efforts have been invested in the experimental measurement of fragment angular momenta (FAM) \cite{Johansson1964,Armbrus1971,Wilhelmy1972,Pleasonton1973,Glassel1989,Travar2021PLB}, culminating with an impressive advance by Wilson $et$ $al$. in 2021 \cite{Wilson2021Nature}: a sawtooth-like mass dependence of FAM with high precision was measured across three different fission reactions. Theoretical studies based on statistical models \cite{Randrup2014PRC,Vogt2017PRC,Vogt2021PRC,Randrup2021PRL,Randrup2022PRC,Randrup2022PRCL,Randrup2024PRC}, microscopic angular momentum projection \cite{Bertsch2019PRC,Marevic2021PRC,Bulgac2021PRL,Bulgac2022PRL,Bulgac2022PRC}, and collective Hamiltonian \cite{Scamps2023PRC,scamps2023PRC-2,scamps2024PRC} can essentially reproduce the distributions of FAM. However, the crucial and challenging issues of how FAM develop dynamically from equilibrium and how the shape fluctuations affect the FAM throughout the entire fission process are still not understood. Clearly, this requires a global, microscopic, and dynamical study to describe the generation and evolution of FAM and includes descriptions of as many quantum fluctuations as possible.

\begin{figure}[t]
    \centering
    {\includegraphics[scale=0.40]{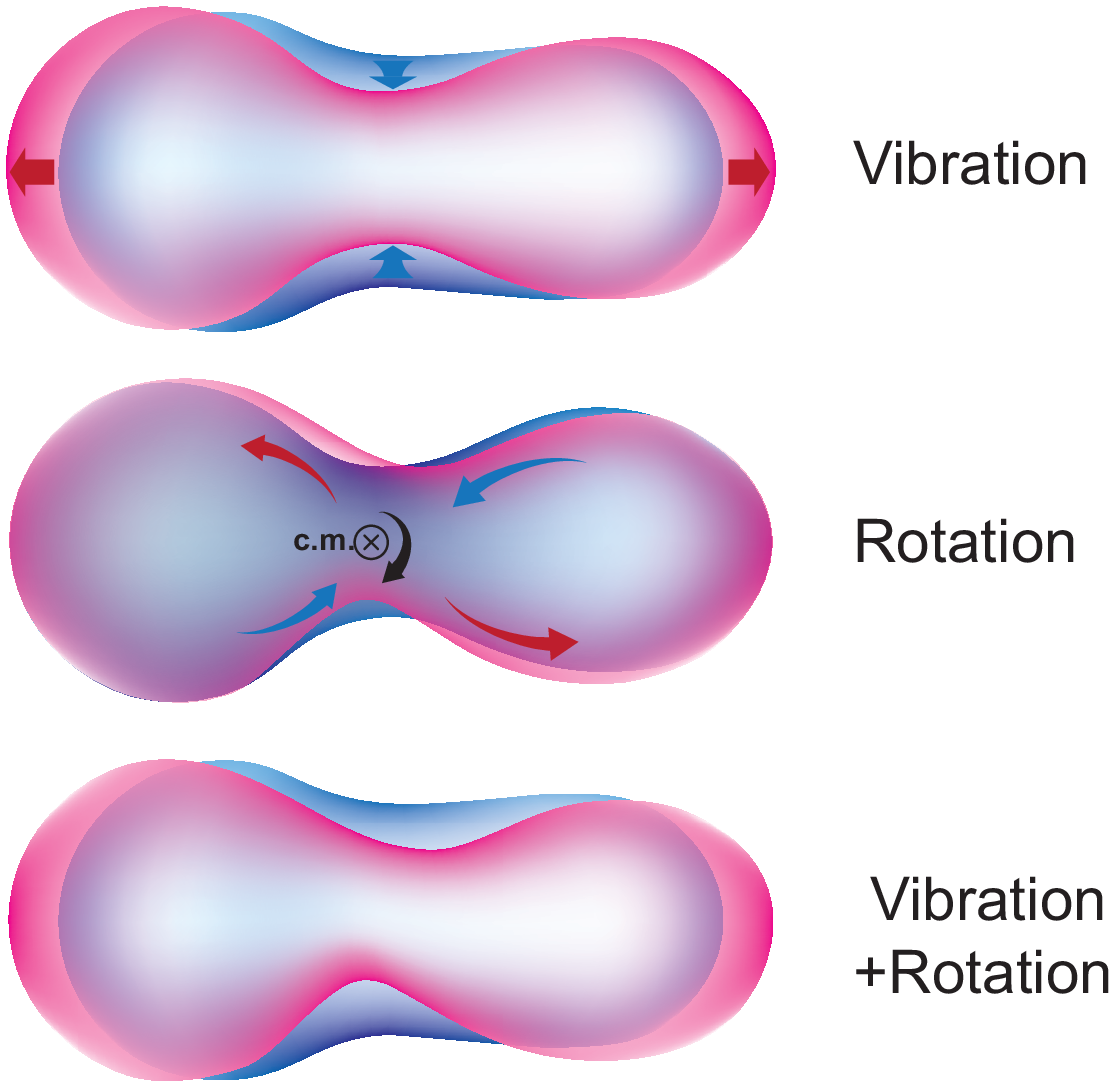}}
    \fontsize{11.5pt}{14.4pt}\selectfont
    \caption{
        Scheme of the collective vibrations (a), rotations (b), and their combinations (c) of a fissioning nucleus caused by quantum fluctuations.}
    \label{fig:model}
\end{figure}

The quantum shape fluctuations in low-energy nuclear fission mainly manifest as collective shape vibrations \cite{Bohrbook}. They can also induce collective rotations, especially when the nucleus goes beyond the saddle point where the elongation of the compound nucleus accelerates. As the fragments are forming \cite{zhangchunli2016}, the collective rotations will gradually converge to those of the two fragments and the relative rotation between them. Figure \ref{fig:model} shows a schematic picture of the collective vibrations, rotations, and their combinations within a fissioning nucleus. These can be treated as fluctuations of the mass multipole moments (shapes of compound nuclei and fragments). These collective rotations are coupled to each other by the multipole-multipole interactions, with the dominant part being the quadrupole-quadrupole interaction \cite{Bohrbook}. 

Therefore, a collective Hamiltonian that includes not only vibrations in a set of collective shape coordinates $\{\boldsymbol{\beta}\}$ but also the rotations of two (forming) fragments and their relative motion can be written as:
\begin{equation}
	\begin{aligned}
	 \hat H=&-\frac{\hbar^{2}}{2} \sum_{p,q}\frac{\partial}{\partial \beta_{p}}B^{-1}_{pq}\frac{\partial}{\partial \beta_{q}}+\sum_{i}\frac{{\hat S}_{i}^{2}}{2 {\cal I}_{i}}+\sum_{i<j}\chi_{2}\hat{Q}^{i}_{2}\cdot \hat{Q}^{j}_{2}\\
	 &+V(\boldsymbol{\beta}),
	 \ \ \ \ \ \ \ \ \ \ \ \ \ \  \ \ i,j=(L,\ H,\ \Lambda)
	\end{aligned}
\end{equation}
where $B_{pq}$ and $V$ are the collective masses and potential, respectively. $S_{L(H)}$ is the angular momentum of the light (heavy) fragment and $S_{\Lambda}$ is the relative orbital angular momentum between two fragments.  ${\cal I}_{L(H)}$ and ${\cal I}_{\Lambda}$ are the corresponding moments of inertia. $\chi_2$ in the quadrupole-quadrupole interaction terms denotes the interaction strength. $\hat Q^i_2$ is the quadrupole operator that is related to the intrinsic quadrupole deformation $\beta^i_2$ \cite{Ring1980Book-P25}.

Then, we can simulate the fission dynamics and, in particular, the generation and continuous evolution of FAM by performing the time evolution of a collective wave function, which takes the form 
\begin{equation}
	\begin{aligned}
		\sum_{n}\phi_{n}(\boldsymbol{\beta},t)\vert S_{L}S_H(S_{LH})S_{\Lambda};JM\rangle	
	\end{aligned}
\end{equation}
 $\phi_{n}(\boldsymbol{\beta},t)$ is the wave function of collective shape coordinates and time $t$ for channel $n$. $n$ denotes the specific coupling of angular momenta of the light and heavy fragments, as well as their relative motion $(S_{L}, S_{H}, S_{\Lambda})$, to the total angular momentum $J$. Here, $J$ is set to zero following the previous works \cite{Randrup2022PRC,Scamps2022PRC,Scamps2023PRC,Randrup2022PRCL}. Finally, we obtain a time-dependent coupled channel equation: 
\begin{widetext}
 \begin{equation}
	 {\rm i}\hbar \frac{\partial}{\partial t} \phi_{n}(\boldsymbol{\beta}, t)=\left[-\frac{\hbar^{2}}{2}\sum_{p,q}\frac{\partial}{\partial \beta_{p}}B^{-1}_{pq}\frac{\partial}{\partial \beta_{q}}+\sum\limits_i\frac{S_{i}(S_{i}+1)\hbar^2}{2 {\cal I}_{i}}+V(\boldsymbol{\beta})\right]\phi_{n}(\boldsymbol{\beta}, t)+\sum_{i<j}\sum_{n^{\prime}\neq n}\Gamma^{ij}_{nn^\prime}\phi_{n^{\prime}}(\boldsymbol{\beta}, t)
    \label{eq:TDMRV}
\end{equation}
\end{widetext}
with
\begin{equation}
	\begin{aligned}
	\Gamma^{ij}_{nn^\prime}=&\chi_2 (-1)^{S_{i}^{\prime}+S_{j}+S_{k}^{\prime}}	
	    \left\{\begin{array}{lll}
		S_{i}^{\prime} & 2 & S_{i} \\
		S_{j} & S_{k}^{\prime} & S_{j}^{\prime}
		\end{array}
	    \right\}\langle S_{i}||\hat Q_{2}^{i}|| S_{i}^{\prime} \rangle\\
        &\times\langle S_{j}||\hat Q_{2}^{j}||S_{j}^{\prime}\rangle\delta_{S_{k} S_{k}^{\prime}}, \ \ \ \ \ \ \ \  (k\neq i\neq j).
	\end{aligned}
\end{equation}

The dynamics of Eq. (\ref{eq:TDMRV}) is fully determined by the shape-dependent collective parameters: $B_{pq}$, ${\cal I}_i$, $V$, and $\beta_2^i$, and the quadrupole-quadrupole interaction strength $\chi_2$. Here, the collective parameters are calculated microscopically with the constrained CDFT \cite{LiZY2024PRC}. It is noteworthy that the moments of inertia of fragments ${\cal I}_{L, H}$ are calculated using the Inglis-Beliaev formula \cite{Inglis1956,Beliaev1961} based on the intrinsic mean-field potential of the fragment, which is deduced from its density distribution under the same CDFT framework ({\it c.f.} Fig. S1 \cite{Supplementmaterial}). In principle, the quadrupole-quadrupole interaction can also be determined approximately from the microscopic calculation \cite{Scamps2022PRC}. Here, to more clearly demonstrate the influence of quadrupole-quadrupole interactions on the generation and evolution of FAM, we use a Fermi distribution for $\chi_{2}$ as a function of $\boldsymbol{\beta}$, which remains almost a constant $c$ from saddle to scission and then decays by half when the two fragments are separated $\approx$ 0.5 fm \cite{Scamps2022PRC}. $c$ is determined by reproducing the pattern of experimental angular momentum distributions for a certain pair of fragments \cite{Wilson2021Nature}. $c = 0.11 {\rm MeV}/b^2$ is used in this work. 

To cpature the generation and continuous evolution of FAM from the equilibrium state of the compound nucleus to a large separation of fission fragments, here we reduce the multiple dimensional collective space to a certain path with the variable $\beta$ ($\delta\beta=\sqrt{\sum_{p} \delta\beta_p^2}$) due to a fact that the collective currents mainly flow over the asymmetric fission valley in the actinides \cite{scampsNature2018,Bulgac2019PRC,chen2023CPC,liFP2024}. Eq. (\ref{eq:TDMRV}) is solved numerically based on a finite difference method in a large-scale collective shape space $\beta$ and the Crank-Nicolson scheme for time integration \cite{Goutte2005}. The angular momenta obey triangular rule $\Delta(S_H S_L S_\Lambda)$, and they are truncated at $36\ \hbar$ for $S_H$, and $40\ \hbar$ for $S_L$ and $S_\Lambda$. When fragments separate far enough, which we set to be the distance between the centers of mass $d_{\rm c.m.}^{\rm FFs} > 23$ fm, they are fully decoupled \cite{Bulgac2021PRL}, and one can calculate the probability flux for each angular momentum channel:
\begin{equation}
    F_{n}(\beta, t) = \int_{t_0}^{t} \boldsymbol{J}_{n}(\beta, t)dt,
    \label{eq:amflux}
\end{equation}
with the current
\begin{equation}
	\begin{aligned}
    & \boldsymbol{J}_{n}(\beta, t) \\
    &= \frac{\hbar}{2{\rm i}} \frac{1}{B(\beta)}  [\phi^{\ast}_{n}(\beta, t) \nabla \phi_{n}(\beta, t) - \phi_{n}(\beta, t) \nabla \phi^{\ast}_{n}(\beta, t)].
	\end{aligned}
\end{equation}

Finally, the two- and one-dimensional probability distributions of FAM can be obtained by summing over the flux with respect to other component(s):
\begin{equation}
    \resizebox{0.98\hsize}{!}{$P(S_L, S_H) = \sum\limits_{S_\Lambda} F_{n}(\beta, t), \ P(S_L\ {\rm or}\ S_H) = \sum\limits_{S_H\ {\rm or}\ S_L} P(S_L, S_H)$}.       
\end{equation}
\vspace{0.15cm}

\begin{figure}[t]
    \includegraphics[scale=0.48]{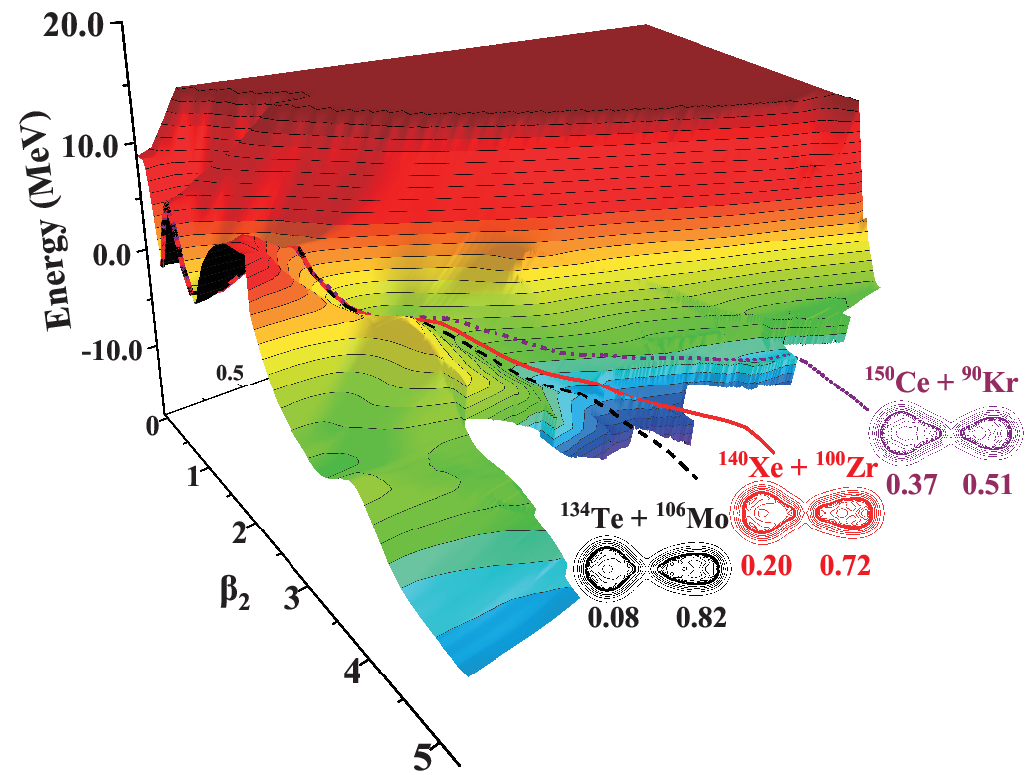}
    \caption{
		The potential energy surface of $^{240}\text{Pu}$ in the $\beta_2-\beta_3$ space calculated by CDFT with the PC-PK1 functional. The contour line is 1 MeV, and the potential energy is set to 0 MeV at the ground state. The fission channels: $^{134}\text{Te} + ^{106}\text{Mo}$, $^{140}\text{Xe} + ^{100}\text{Zr}$, and $^{150}\text{Ce} + ^{90}\text{Kr}$ are shown by the dashed, solid, and dotted lines, respectively. The corresponding scission configurations and quadrupole deformations of the nascent fission fragments are also shown.
		}
        \vspace{-0.3cm}
    \label{fig:pes}
\end{figure}

In this work, we take the fission of the compound nucleus $^{240}$Pu as an example to investigate the generation and continuous evolution of FAM and analyze the impact of quantum shape fluctuations. Figure \ref{fig:pes} displays the potential energy surface (PES) in the quadrupole ($\beta_2$) and octupole ($\beta_3$) deformation space calculated by CDFT \cite{LiZY2024PRC} with the PC-PK1 functional \cite{ZhaoPW2010}. The inner and outer fission barriers and asymmetric fission valleys out of ($\beta_2, \beta_3$) $\approx$ (1.3, 0.5) are clearly identified, which are consistent with previous calculations using Skyrme and Gogny density functionals \cite{MarevicP2020,ChenYJ2022,Younes2012,Regnier2016PRC}. Based on the PES, the low-energy fission dynamical simulation using the evolution of collective spatial wave functions \cite{Regnier2018} shows that the probability currents mainly follow the asymmetric fission valley and its surroundings ({\it c.f.} Fig. S2 and video \cite{Supplementmaterial}). Therefore,  We choose three fission channels in the asymmetric fission valley: $^{134}$Te+$^{106}$Mo, $^{140}$Xe+$^{100}$Zr, and $^{150}$Ce+$^{90}$Kr as examples to perform the simulations, where six fragments spread across the region of the sawtooth-like distribution of FAM \cite{Wilson2021Nature}. The fragment information at scission is also shown in Fig. \ref{fig:pes} and we find that the quadrupole deformations of the heavy (light) fragments increase (decrease) gradually as the mass asymmetry increases.

\begin{figure}[b]  
    \includegraphics[scale=0.48]{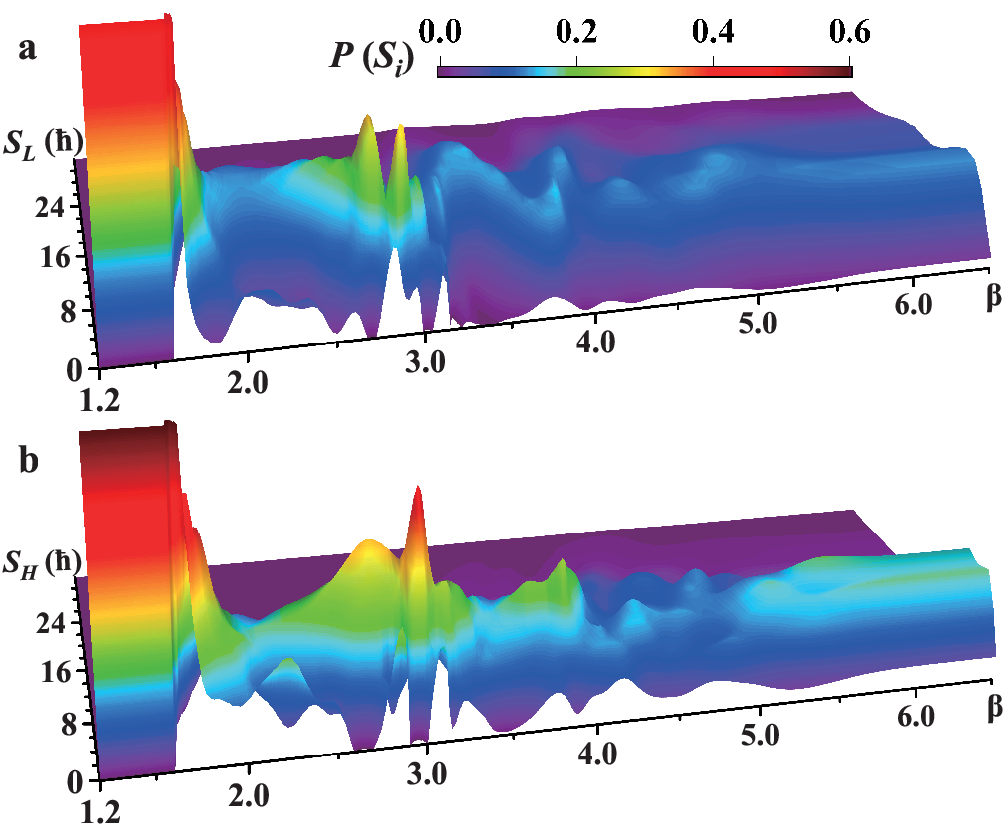}
    \selectfont\caption{
        Angular momentum distributions for the light (upper) and heavy (lower) fragments along the fission path for the channel $^{140}\text{Xe} + ^{100}\text{Zr}$.}
        \vspace{-0.3cm}
    \label{fig:AMD-beta}
\end{figure}
With all the microscopic inputs--potential energy curves, collective masses, moments of inertia and quadrupole deformations of the fragments ({\it c.f.} Fig. S3 -- S6 \cite{Supplementmaterial} )--we can simulate The dynamics of FAM using Eq. (\ref{eq:TDMRV}). Figure \ref{fig:AMD-beta} displays the angular momentum distributions for the fragments along the fission path for the channel $^{140}$Xe+$^{100}$Zr as an example. The corresponding two-dimensional (2D) FAM distributions can be found in the videos \cite{Supplementmaterial}. Remarkably, during fission fragment formation, i.e., from the saddle point ($\beta \approx 1.7$) to scission ($\beta \approx 5.2$), we observe a rapid generation and chaotic evolution of the FAM distribution, which are driven by the shape fluctuations induced mainly by the sharply descending potential energy and quadrupole-quadrupole interactions. After scission, the distribution varies slightly till to a stable pattern when the fragments are well separated. Moreover, we find that the evolution of angular momentum distributions for $S_L$, $S_H$, and $S_\Lambda$ have similar features ({\it c.f.} Fig. S7 \cite{Supplementmaterial}). Here, we present a novel depiction of the generation and evolution of FAM, which offers a fresh perspective on the dynamics of angular momentum in nuclear fission.

\begin{figure}[h]  
    \includegraphics[scale=0.398]{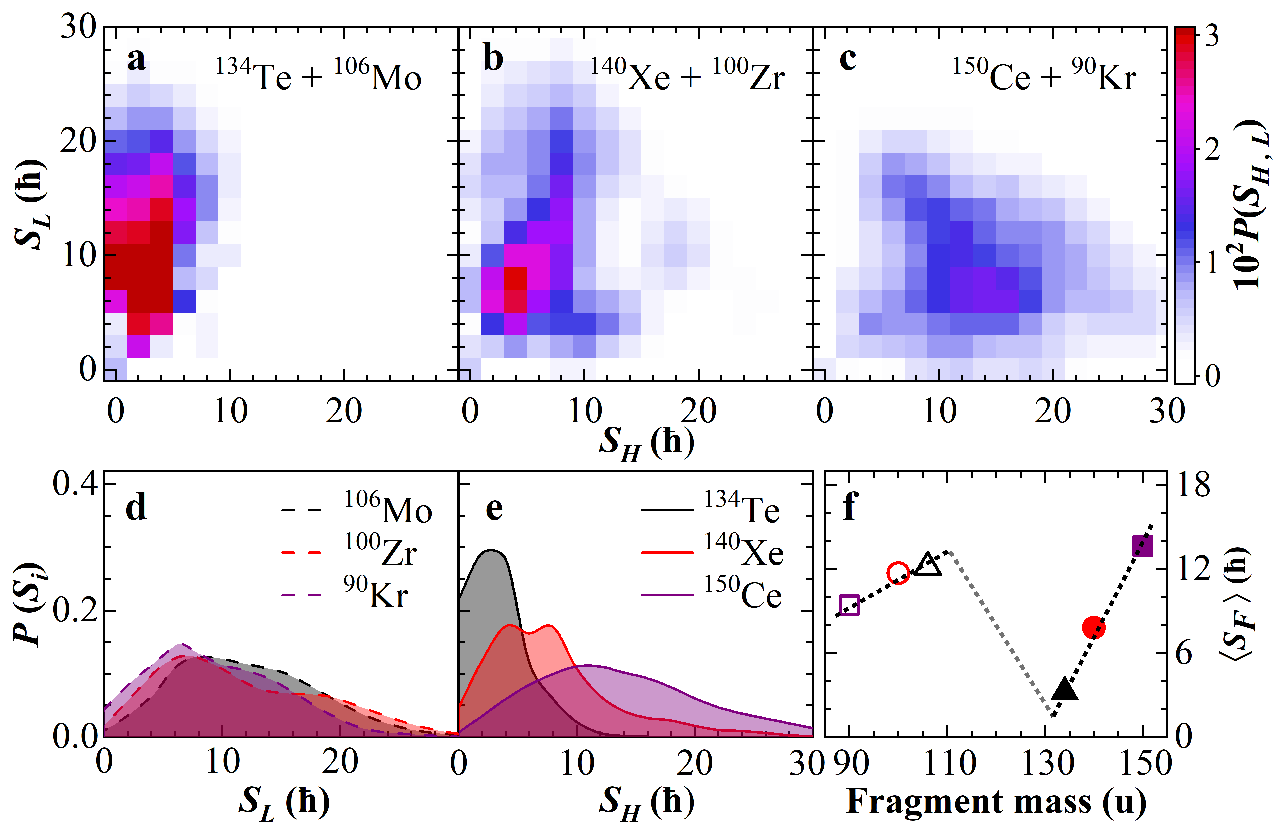}
    \selectfont\caption{
        The 2D (a-c) and 1D (d-e) probability distributions of FAM for three fission channels of $^{240}\text{Pu}$. (f) Mass dependence of average angular momenta for six fragments. The dashed lines are used to guide eyes.}
    \label{fig:AMD-distribution}
\end{figure}


Figure \ref{fig:AMD-distribution} displays the final FAM distributions and the mass dependence of the average FAM. Extended peaks are observed in the 2D distributions (Fig. \ref{fig:AMD-distribution}(a-c)) for all three fission channels, and the peaks and entire patterns move toward larger $S_H$ (and lower $S_L$) as the mass asymmetry of the fragments increases. The corresponding one-dimensional (Fig. \ref{fig:AMD-distribution}(d), (e)) distributions show this trend in projection: the peak for the heavy fragment moves from $2\hbar$ ($^{134}$Te) to $10\hbar$ ($^{150}$Ce) and the width extends dramatically, which is consistent with the measurements in the Extended Data of Fig. 4 of Ref. \cite{Wilson2021Nature}. Consequently, the mass dependence of average angular momenta for the six fragments presents a clear sawtooth-like pattern (Fig. \ref{fig:AMD-distribution}(f)), which also implies positive correlations with moments of inertia of the nascent fragments ({\it c.f.} Fig. S4 \cite{Supplementmaterial}) and the quadrupole deformations ({\it c.f.} Fig. S5 \cite{Supplementmaterial}).

\begin{figure}[h]  
    \includegraphics[scale=0.57]{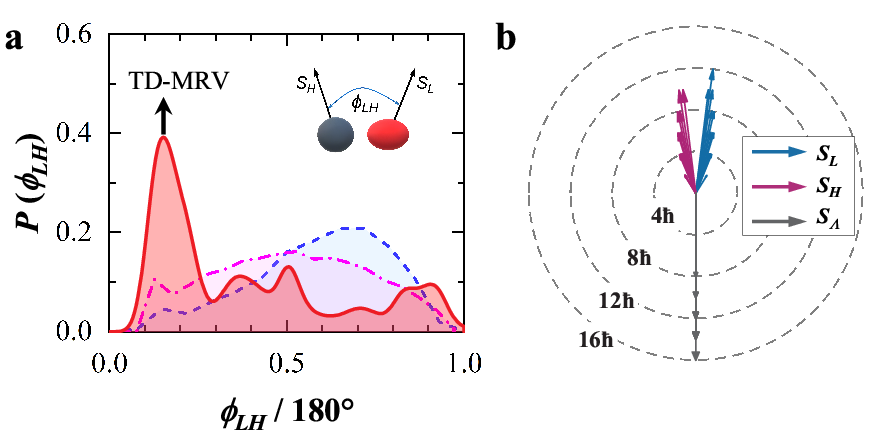}
    \selectfont\caption{
        (a) The opening angle distributions smoothed by Gaussian convolution with $\sigma = 5^\circ$ (solid line) for the fission channel $^{140}\text{Xe} + ^{100}\text{Zr}$ calculated by TD-MRV. The results from angular momentum projection (dashed line) and modified FREYA (dash-dotted line) are also shown. (b) The angular momentum geometry around the peak with an opening angle $\phi_{{LH}} \approx 30^\circ$.}
    \vspace{0cm}
    \label{fig:AMD-corrletion}
\end{figure}

The correlations of FAM orientations are analyzed in Fig. \ref{fig:AMD-corrletion}(a). It is remarkable to find that the distribution of FAM orientations described by the opening angle $\phi_{LH}$ \cite{Randrup2021PRL,Scamps2023PRC} displays strong correlations at the small, medium, and large opening angles ($\phi_{LH} \approx 30^\circ, 90^\circ, 160^\circ$). Furthermore, we find that the peak at $\phi_{LH} \approx 30^\circ$ is dominated by the wriggling mode \cite{NIX19651} as shown in Fig. \ref{fig:AMD-corrletion}(b). The correlations differ from those predicted by the angular momentum projection \cite{Bulgac2022PRL} and statistical model (modified FREYA) \cite{Bulgac2022PRC}, where broad peaks $\phi_{LH} \approx 120^\circ$ and $90^\circ$ are predicted. We note that the quantum fluctuations were not considered in those models. Such anisotropic correlations of FAM can manifest themselves in the correlation between $\gamma$ transitions emitted from two fragments, as reported in Ref. \cite{Smith1999}, where a hint for the enhancement of near-aligned or antialigned $\gamma$-$\gamma$ emission was observed. Further experimental studies are expected to provide quantitative evaluations of the calculated opening angles.

\vspace{0.15cm}
In conclusion, we have investigated the generation and evolution of FAM from the equilibrium of the compound nucleus to a large separation of fission fragments in a microscopic framework that incorporates critical quantum shape fluctuations for the first time. The probability distributions of FAM, as calculated, align well with experimental observations, accurately reproducing the characteristic sawtooth-like mass dependence of average FAM. Our findings emphasize the uniqueness of quantum shape fluctuations in the generation, chaotic evolution, and strong correlations of FAM. Future experiments focusing on the correlation between $\gamma$ transitions from fission fragments are anticipated to provide quantitative insights into the correlations of FAM orientations, thereby substantiating the influence of quantum shape fluctuations. Our work not only deepens the fundamental understanding of nuclear fission dynamics but also has significant implications for the $\gamma$-ray heating problem in nuclear reactors and the synthesis of superheavy elements. Moreover, this work could inspire novel methodologies in studying complex dynamic processes of many-body systems in various scientific domains, ranging from atomic physics to cosmology, where quantum fluctuations and angular momentum are integral to understanding the underlying dynamics.

\vspace{0.15cm}
\textit{Acknowledgements}.---Fruitful discussions with Dr. Song Guo, Dr. Longjun Wang, and Dr. Sibo Wang are gratefully acknowledged. This work was partly supported by the National Natural Science Foundation of China (Grants No. 12375126), the Fundamental Research Funds for the Central Universities, and the U.S. Dept. of Energy, Office of Science, Office of Nuclear Physics under Contract No. DE-AC05- 00OR22725 (ORNL).

\bibliography{apssamp}


\setcounter{figure}{0}
\renewcommand{\thefigure}{S\arabic{figure}}

\vspace{5cm}
\begin{table*}
{\fontsize{30}{48} \selectfont \textbf{Supplemental Materials}}
\end{table*}

\begin{figure*}[b]
    \centering
    \includegraphics[scale=0.95]{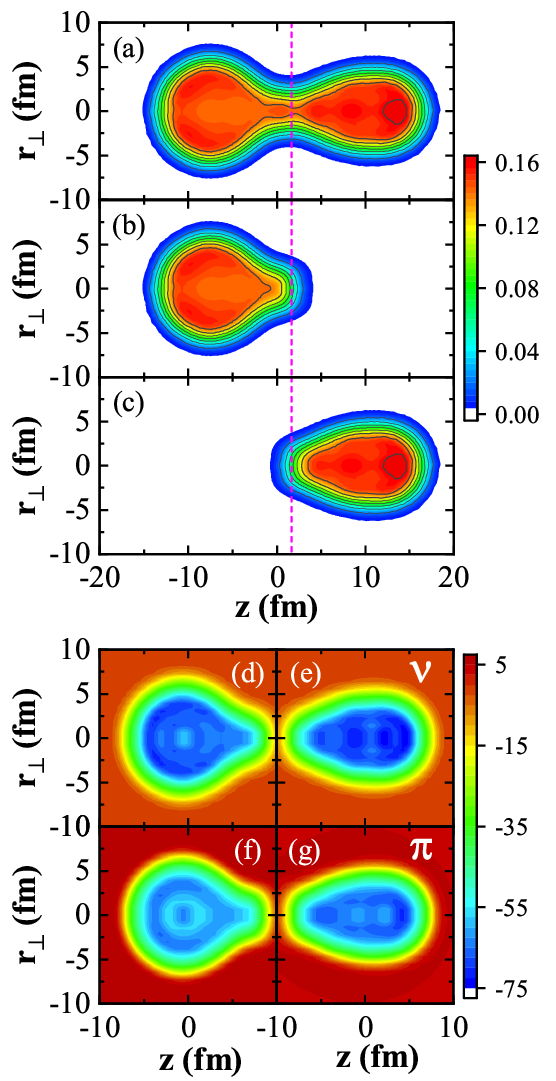}
    \caption{(a) The nucleon density distribution at $(\beta_2,\ \beta_3)=(3.47,\ 1.91)$ of fissioning compound nucleus $^{240}$Pu. (b) The density distributions for the left fission fragment extracted from that of compound nucleus using a Fermi distribution {\large $f(z)=\frac{1}{1+\exp[(z-z_0)/a]}$} with the neck position $z_0$ and diffuseness $a=0.67$ fm. (c) Same as panel (b) but for the right fragment with a Fermi distribution $1-f(z)$. (d-g) The neutron and proton intrinsic mean-field potentials for fission fragments determined by the corresponding densities using the same energy density functional as the whole nucleus.}
    \label{figs1:ffstruct}
\end{figure*}

\begin{figure*}
    \centering
    \includegraphics[scale=0.8]{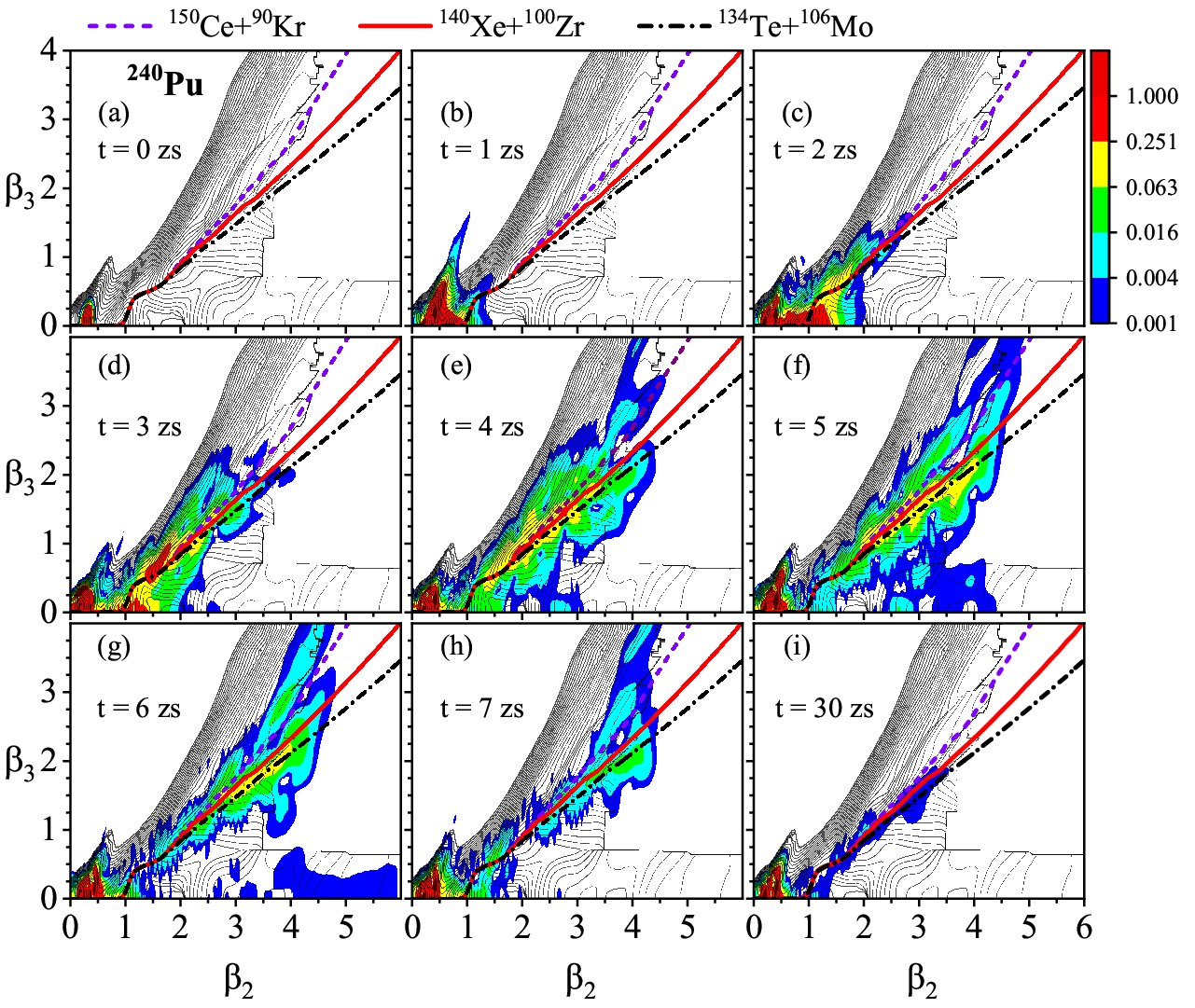}
    \caption{Evolution of collective probability density distribution in the $\beta_2-\beta_3$ plane for $^{240}$Pu simulated by the time-dependent generator coordinate method \cite{Regnier2018}.}
    \label{figs2:wvf-evo}
\end{figure*}

\begin{figure*}
    \includegraphics[scale=0.8]{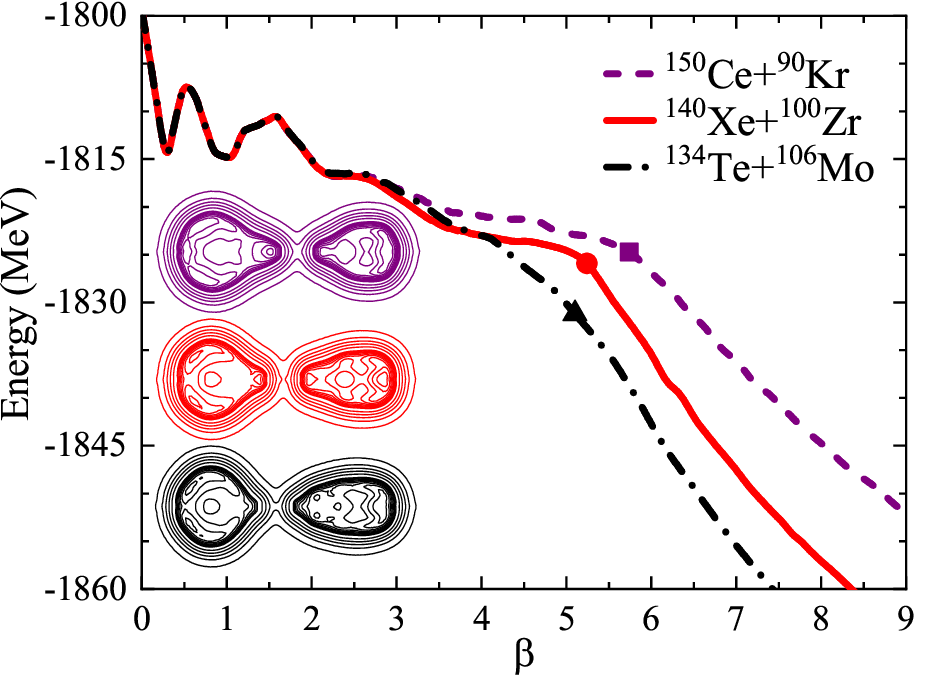}
   \caption{The potential energy curves and scission points (solid symbols) along the $\beta$ path for the three fission channels. The scission configurations are determined in the $(\beta_2, \beta_3, q_N)$ space with the nucleon number in the neck $q_N$ = 1. Configurations beyond scission are achieved by imposing constraint on the scission density distribution but splitting the two fragments by a certain distance.}
   \label{figs3:PECs}
\end{figure*}

\begin{figure*}
    \includegraphics[scale=0.8]{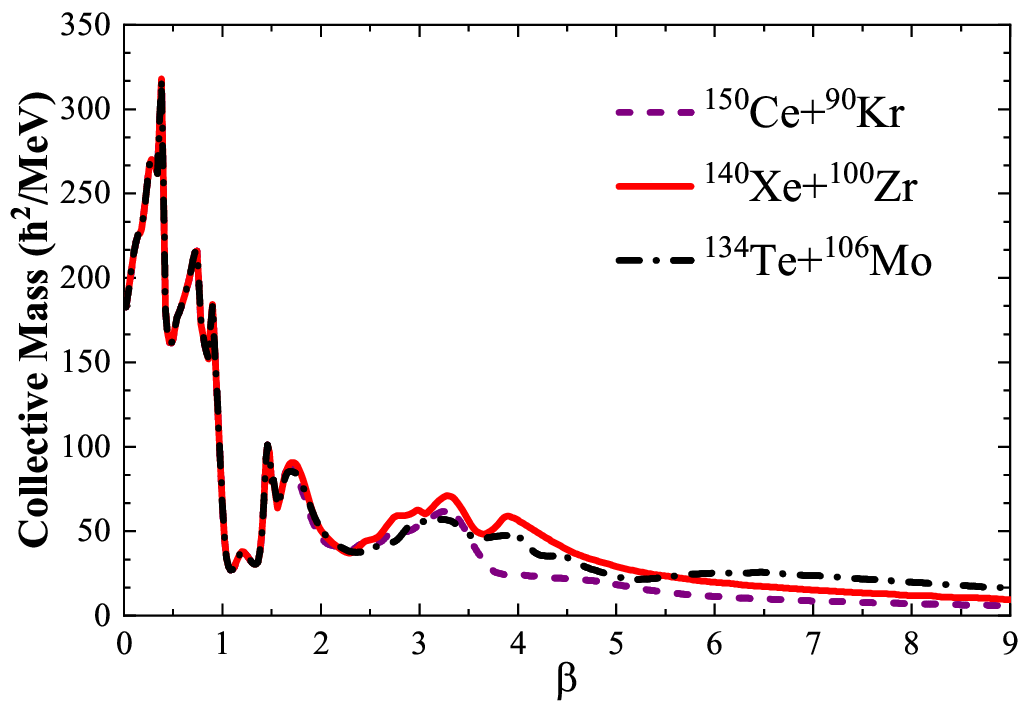}
    \caption{The collective masses for the three fission channels as functions of $\beta$.}
    \label{figs4:collmass}
\end{figure*}

\begin{figure*}[ht]
    \centering
    \begin{minipage}[t]{0.48\linewidth}
        \includegraphics[scale=0.99]{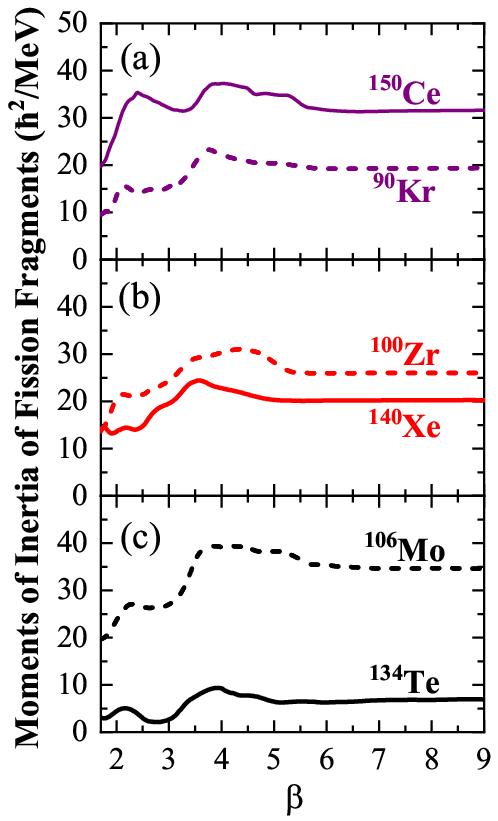}
        \caption{Moments of inertia of fission fragments as functions of  $\beta$.}
        \label{figs5:ffinermnt}
    \end{minipage}
    \hfill\hfill
    \begin{minipage}[t]{0.48\linewidth}
        \includegraphics[scale=0.99]{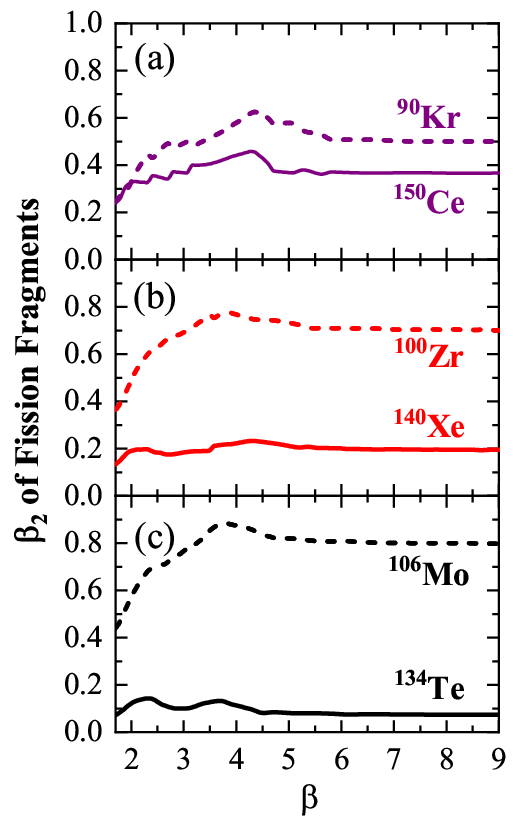}
        \caption{Quadrupole deformations $\beta_2$ of fission fragments as functions of $\beta$.}
        \label{figs6:ffbeta}
    \end{minipage}

\end{figure*}

\begin{figure*}
    \includegraphics[scale=1.1]{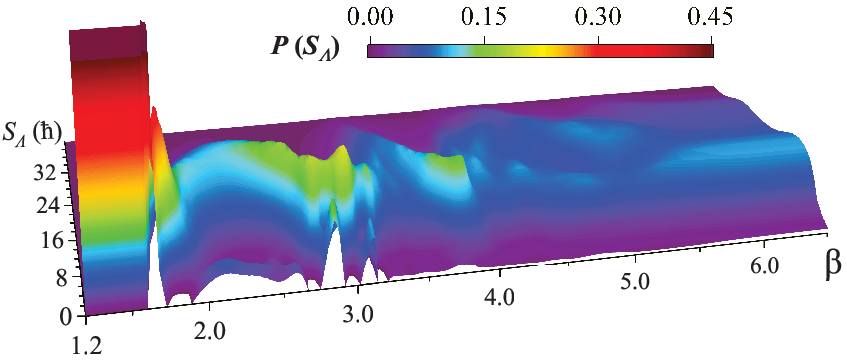}
   \caption{Angular momentum distributions for the relative rotation between two fragments along the fission path for the channel $\prescript{140}{}{\rm Xe} + \prescript{100}{}{\rm Zr}$.}
   \label{figs7:P-beta-Lamda}
\end{figure*}

\begin{figure*}
    \includegraphics[scale=0.8]{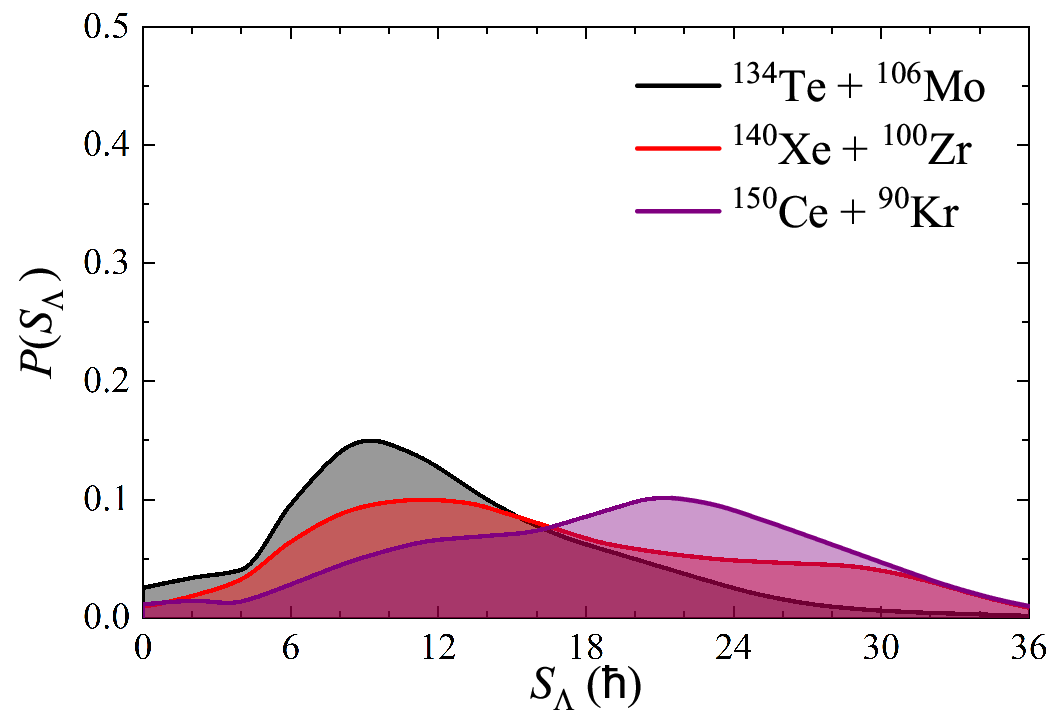}
   \caption{The final probability distributions for the relative rotations between the partner fragments.}
   \label{figs8:Lamda}
\end{figure*}

\end{document}